\documentclass[12pt]{article} \textheight 8.5in \textwidth 6.5in
\usepackage{amsmath,amsfonts,latexsym,amssymb}

\newcommand{\Ro}{\tilde R}
\newcommand{\nablao}{\tilde \nabla}
\newcommand{\tgo}{\tilde g}

\oddsidemargin 0in \topmargin -.35in \title{Conformal Null Infinity
Does Not Exist for Radiating Solutions in Odd Spacetime Dimensions} 
\author{Stefan Hollands\thanks{\tt stefan@gr.uchicago.edu} \,\, and
Robert M. Wald\thanks{\tt rmwa@midway.uchicago.edu} \\ \it Enrico
Fermi Institute and Department of Physics \\ \it University of Chicago
\\ \it 5640 S.~Ellis Avenue, Chicago, IL~60637, USA}
\begin{document}

\maketitle
\begin{abstract}

We show that for general relativity in odd spacetime dimensions
greater than $4$, all components of the unphysical Weyl tensor for
arbitrary smooth, compact spatial support perturbations of Minkowski
spacetime fail to be smooth at null infinity at leading nonvanishing
order. This implies that for nearly flat radiating spacetimes, the
non-smoothness of the unphysical metric at null infinity manifests
itself at the same order as it describes deviations from flatness of
the physical metric. Therefore, in odd spacetime dimensions, it does
not appear that conformal null infinity can be in any way useful
for describing radiation.

\end{abstract}

The notion of conformal null infinity was introduced more than forty
years ago by Penrose \cite{p} in the context of $4$-dimensional
general relativity, and has provided a remarkably fruitful framework for
giving a mathematically precise description of the asymptotic
properties of gravitational radiation. In this framework, one defines
an ``unphysical spacetime'' with a metric that is conformally related
to the physical metric, in such a way that ``asymptotically large
distances in null directions'' correspond to ordinary points on a
boundary (``null infinity'') attached to the unphysical spacetime.
One can then define notions such as Bondi mass and energy flux via
local definitions on this boundary rather than by taking asymptotic
limits in the original, physical spacetime. The appropriate degree of
differentiability to assume for the unphysical metric has been been a
subject of much discussion and analysis (see \cite{f} and references
cited therein). However, in $4$-dimensional general relativity
it is known that there is a wide class of
radiating spacetimes for which the unphysical metric is smooth
\cite{cd}.

The notion of conformal null infinity and the definitions of Bondi
mass and energy flux have recently been investigated by Hollands and
Ishibashi \cite{hi} in the context of general relativity in spacetimes
of dimension greater than $4$. It was found in \cite{hi} that for even
dimensional spacetimes, the notion of asymptotic flatness can be
generalized in such a way that notions of Bondi mass and energy flux
can be defined. It also was found that this definition of asymptotic
flatness is stable to small perturbations in the sense that linearized
perturbations with smooth initial data of compact support off of an
asymptotically flat solution satisfy the linearization of the
asymptotic flatness criteria. (This property had previously been shown
to hold in $4$-dimensions by Geroch and Xanthopoulos \cite{gx}.) However,
Hollands and Ishibashi also noted that their proposed definition of
asymptotic flatness would not work in odd dimensions, since, in their
proposal, in odd dimensions the leading order deviation of the unphysical
metric from a fixed smooth background metric would be proportional to
a half-integral power of the conformal factor $\Omega$. Consequently, the
unphysical metric would not have sufficient smoothness to perform any
meaningful constructions at null infinity.

The main purpose of this paper is to show that these difficulties are
not merely artifacts of the particular proposed definition in
\cite{hi}. In odd spacetime dimensions, there is no difficulty in
conformally mapping Minkowski spacetime into a portion of the Einstein
static universe (see below). (Similarly, there is no difficulty in
defining a suitable conformal null infinity structure in other
stationary spacetimes\footnote{It should be noted that in $4$
spacetime dimensions, the deviation of the physical metric from the
Minkowski metric will fall off at null infinity as $1/r$ for both
stationary and radiating solutions. However, in both even and odd
dimensional spacetimes with $d>4$ the deviation of a stationary metric
from the Minkowski metric will have a faster fall-off rate than that
of a generic asymptotically flat radiating solution~\cite{hi}.}, such
as odd dimensional Schwarzschild spacetime). However, severe
difficulties with conformal null infinity in odd spacetime dimensions
arise when one considers spacetimes containing
radiation\footnote{The fact that Huygens' principle fails for the
wave equation in odd dimensional Minkowski spacetime suggests the
possibility that the failure of Huygens' principle might be
responsible for the difficulties with the conformal null infinity
framework. However, our analysis does not reveal any obvious direct
connection between the failure of Huygens' principle and the failure
of the conformal null infinity framework}, and these difficuties
manifest themselves already at the level of linearized perturbations
off of Minkowski spacetime. In this paper, we will show that for
perturbations of odd dimensional Minkowski spacetime with initial data
that is smooth and of compact support, the fall-off of the physical
Weyl tensor is such that the behavior of the unphysical Weyl tensor at
null infinity must begin with a helf-integral power of $\Omega$. This
shows not only that the perturbed unphysical metric cannot be smooth
at null infinity, but furthermore that its failure to be smooth
manifests itself at the same order as it describes deviations of the
physical metric from flatness. It therefore does not seem possible
that useful constructions to describe gravitational radiation can be
given within a conformal null infinity framework in odd dimensional
spacetimes.

To begin, consider $d$-dimensional Minkowski spacetime $(M \equiv {\mathbb R}^d,
\eta_{ab})$ for any (even or odd) $d > 3$, where $\eta_{ab}$ denotes
the flat metric
\begin{eqnarray}
ds^2 &=& -dt^2 + dx_1^2 + \dots + dx_{d-1}^2 \nonumber \\
&=& -dt^2 + dr^2 + r^2 d\omega_{d-2}^2 \, ,
\label{mink}
\end{eqnarray}
where $d\omega_{d-2}^2$ denotes the metric on the unit, round
$(d-2)$-dimensional sphere. The $d$-dimensional Einstein static universe 
is the manifold $E \equiv {\mathbb R} \times S^{d-1}$ with metric $\tgo_{ab}$
given by
\begin{equation}
d \tilde s^2 
= -dT^2 + dR^2 + \sin^2 R \, d\omega_{d-2}^2 \, .
\label{ein}
\end{equation}
As in $4$-dimensions (see, e.g., \cite{w}), we can conformally map
$d$-dimensional Minkowski spacetime into a portion of the
$d$-dimensional Einstein static universe as follows: We define
$u=t-r$, $v=t+r$, and we define the map $\psi: M \rightarrow E$ so as
to preserve the $(d-2)$-sphere angular coordinates and take the point
labeled by $(u,v)$ into the point labeled by
\begin{eqnarray}
T &=& \tan^{-1}v + \tan^{-1} u \,, \\
R &=& \tan^{-1}v - \tan^{-1} u \, .
\end{eqnarray}
Then it can be verified that 
\begin{equation}
\tgo_{ab} = \Omega^2 \psi^* \eta_{ab} \, ,
\end{equation}
where
\begin{equation}
\Omega^2 = 4 (1 + v^2)^{-1}(1 + u^2)^{-1} \, .
\end{equation}
Note that this construction works equally well for $d$ odd or even.

The null portion of the boundary of Minkowski spacetime in the
Einstein static universe---namely, the null surfaces $\mathcal{I}^+$
and $\mathcal{I}^-$, defined, respectively, by $T=\pi - R$ and $T = R -
\pi$, with $0 < R < \pi$---provides the prototype for the construction
of conformal null infinity for asymptotically flat spacetimes.
However, in order for this construction to be useful for describing
spacetimes where gravitational radiation is present, it is necessary
for this construction to continue to work if we perturb Minkowski
spacetime to another solution of the vacuum Einstein equation $R_{ab}
= 0$. This leads to the following requirement at the level of linear
perturbation theory: Let $\gamma_{ab}$ be a solution of the linearized
Einstein equation with suitably regular initial data (say, smooth and
of compact support).  Then, modulo a gauge transformation of
$\gamma_{ab}$ and a perturbation, $\delta \Omega$, of the conformal
factor, we require that the perturbed unphysical metric extend to
conformal null infinity so as to be sufficiently differentiable
there. Here, by ``sufficiently differentiable'', we mean (at the very
least) that some non-Minkowskian asymptotic properties of the physical
metric $\eta_{ab} + \gamma_{ab}$ can be expressed in terms of the
curvature of the unphysical metric (or derivatives of its curvature)
evaluated at conformal null infinity. Thus, given $\gamma_{ab}$, we
seek a vector field $\xi^a$ on Minowski spacetime and a smooth
function $\delta \Omega$ on the Einstein static universe (with $\delta
\Omega = 0$ at conformal null infinity) such that the perturbed
unphysical metric
\begin{equation}
\tilde{\gamma}_{ab} = \Omega^2 \psi^*(\gamma_{ab} + 
\partial_a \xi_b + \partial_b \xi_a)
+ 2\Omega \delta \Omega \psi^* \eta_{ab}
\end{equation}
extends to conformal null infinity so as to be sufficiently
differentiable there. (Here $\partial_a$ denotes the derivative
operator associated with $\eta_{ab}$ and indices are raised and
lowered with $\eta_{ab}$.) If the spacetime dimension $d$ is even, it
has been shown \cite{gx}, \cite{hi} that---by a suitable choice of
$\xi^a$ and $\delta \Omega$---the perturbed unphysical metric
$\tilde{\gamma}_{ab}$ can always be made to be smooth. Indeed, as
already mentioned above, smoothness holds not only for perturbations
of Minkowski spacetime but for perturbations of any asymptotically
flat vacuum solution. However, we now shall show that this is not the
case when $d$ is odd.

Rather than analyze the behavior of linearized perturbations in terms
of the metric perturbation $\gamma_{ab}$, it is much more convenient
to work directly with the perturbed Weyl tensor, $\delta
C_{abcd}$. This has the dual advantage that $\delta C_{abcd}$ is both
gauge invariant and conformally invariant. Consequently, by working
with $\delta C_{abcd}$ we will be able to immediately determine the
behavior of the unphysical Weyl tensor $\delta \tilde{C}_{abcd}$ near
conformal null infinity. We will show that in odd dimensions, the
leading order behavior of $\delta \tilde{C}_{abcd}$ must always begin
with a half-integral power of $\Omega$, and thus it is non-smooth to the
same extent as it is nonvanishing.

First, consider any spacetime that satisfies the vacuum Einstein
equation $R_{ab} = 0$. The uncontracted Bianchi identity immediately
yields
\begin{equation}
\nabla_{[a} C_{bc]de} = 0 \, .
\label{b1}
\end{equation}
Contracting over $a$ and $d$ and using the tracelessness of the Weyl
tensor, we obtain
\begin{equation}
\nabla^{a} C_{bcae} = 0 \, .
\label{b2}
\end{equation}
Applying $\nabla^a$ to eq.(\ref{b1}), we obtain
\begin{equation}
0 = \nabla^a \nabla_{[a} C_{bc]de} = \frac{1}{3} \nabla^a \nabla_{a} C_{bcde}
+ \frac{1}{3} \nabla^a \nabla_b C_{cade} 
+ \frac{1}{3} \nabla^a \nabla_c C_{abde}   \, .
\label{b3}
\end{equation}
But, we also have
\begin{eqnarray}
\nabla^a \nabla_b C_{cade} &=& \nabla_b \nabla^a C_{cade} +
C^a{}_{bc}{}^f C_{fade} + C^a{}_{ba}{}^f C_{cfde} + C^a{}_{bd}{}^f C_{cafe} + 
C^a{}_{be}{}^f C_{cadf} \nonumber \\
&=& C^a{}_{bc}{}^f C_{fade} + C^a{}_{bd}{}^f C_{cafe} + 
C^a{}_{be}{}^f C_{cadf} \, ,
\end{eqnarray}
and similarly for the last term in eq.(\ref{b3}).
Consequently, we obtain
\begin{equation}
\nabla^a \nabla_{a} C_{bcde} = {\rm terms \,\, quadratic
\,\, in}\,\, C \, .
\end{equation}

It follows immediately from the above formulas that for any linearized
perturbation $\gamma_{ab}$ off of Minkowski spacetime that satisfies the
linearized Einstein equation $\delta R_{ab} = 0$, the linearized Weyl
tensor $\delta C_{abcd}$ satisfies
\begin{equation}
\partial^{a} \delta C_{abcd} = 0 \, 
\label{c1}
\end{equation}
and
\begin{equation}
\partial^e \partial_e \delta C_{abcd} = 0\, .
\label{c2}
\end{equation}

For each $\mu = 0,1,\dots d-1$ let $K_\mu^a$ denote the translational
Killing field in Minkowski spacetime assoicated with the $\mu$th
coordinate basis field of the global inertial coordinates $(t=x^0,
x^1, \dots, x^{d-1})$ of eq.~(\ref{mink}), i.e., let $K_\mu^a =
(\partial/\partial x^\mu)^a$.  Then we have $\partial_a K_\mu^b =
0$. Let $W^a$ denote the dilational conformal Killing field of
Minkowski spacetime whose components in the global inertial
coordinates are given by $W^\mu = x^\mu$. Then we have
\begin{equation}
\partial_a W^b = \delta_a{}^b \, .
\label{w}
\end{equation}
For $i = 1,2,3,4$, let
\begin{equation}
V^a_i = \sum_\mu a_i^\mu K_\mu^a + b_i W^a
\label{V}
\end{equation}
where the $a_i^\mu$ and $b_i$ are arbitrary constants.
Let
\begin{equation}
\phi =  \delta C_{abcd}V^a_1 V^b_2 V^c_3 V^d_4  \, .
\end{equation}
Then, we have
\begin{eqnarray}
\partial_e \phi = (\partial_e \delta C_{abcd})V^a_1 V^b_2 V^c_3 V^d_4
&+& b_1 \delta C_{ebcd} V^b_2 V^c_3 V^d_4 + b_2 \delta C_{aecd} V^a_1
V^c_3 V^d_4 \nonumber \\
&+& b_3 \delta C_{abed} V^a_1 V^b_2 V^d_4 + b_4 \delta
C_{abce} V^a_1 V^b_2 V^c_3 \, .
\end{eqnarray}
Applying $\partial^e$ to this equation and using eqs.~(\ref{c1}),
(\ref{c2}), (\ref{w}), and the tracelessness of the Weyl tensor, we obtain
\begin{equation}
\partial^e \partial_e \phi = 0  \, .
\end{equation}
Consequently, on the Einstein static universe, the quantity
\begin{equation}
\tilde{\phi} \equiv \Omega^{1-d/2} \phi
\end{equation}
satisfies the conformally invariant wave equation
\begin{equation}
\nablao  {}^e  \nablao_e  \tilde{\phi} - 
\frac{d-2}{4(d-1)} \Ro  \tilde{\phi} = 0 \, ,
\end{equation}
(see, e.g., appendix D of \cite{w}) where $\Ro$ denotes the
scalar curvature of the Einstein static universe and, in this
equation, indices are raised and lowered with the 
unphysical metric, $\tgo_{ab}$ (see eq.~(\ref{ein})). It then follows
immediately from the well posedness of the initial value formulation
of this equation on the Einstein static universe that $\tilde{\phi}$
extends smoothly to conformal null infinity and cannot vanish
everywhere on $\mathcal{I}^+$ or on $\mathcal{I}^-$ unless $\phi$
itself vanishes identically. Since we have
\begin{equation}
\delta \tilde{C}_{abcd} = \Omega^2 \delta C_{abcd}
\end{equation}
(where $\delta \tilde{C}_{abcd}$ denotes the perturbed unphysical Weyl
tensor with its index lowered by $\tgo_{ab}$) we have learned that
for any perturbation of Minkowski spacetime with smooth initial data
of compact support, and for any choice of the constants $a_i^\mu$ and $b_i$
(see eq.(\ref{V}) above), the quantity
\begin{equation}
\chi \equiv \Omega^{-1-d/2} \delta \tilde{C}_{abcd} V^a_1 V^b_2 V^c_3 V^d_4
\label{chi0}
\end{equation}
is smooth everywhere in the Einstein static universe, and it cannot vanish
everywhere on $\mathcal{I}^+$ (or on $\mathcal{I}^-$) unless $\chi$ vanishes
identically.

By direct calculation, it can be verified that in the Einstein static
universe, we have
\begin{equation}
K_\mu^a =  \alpha_\mu n^a - \Omega  \nablao  {}^a \alpha_\mu  \, .
\label{K}
\end{equation}
Here $n^a \equiv \nablao  {}^a \Omega$, each $\alpha_\mu$ is smooth, and
again indices are raised and lowered with $\tgo_{ab}$. Since $\alpha_0$
is nonvanishing in a neighborhood of $\mathcal{I}^+$, we may substitute for
$n^a$ in terms of $t^a = K_0^a$ and rewrite this
equation as
\begin{equation}
K_\mu^a =  \beta_\mu t^a - \alpha_0 \Omega  \nablao  {}^a \beta_\mu  \, ,
\label{K2}
\end{equation}
where $\beta_\mu = \alpha_\mu/\alpha_0$, so, in particular, we have
$\beta_0 = 1$. For $\mu = 1, \dots, d-1$ we define $s_\mu^a = \nablao
{}^a \beta_\mu$. Then it is not difficult to verify that on
$\mathcal{I}^+$ the vector fields $t^a, s_1^a, \dots, s_{d-1}^a$
span\footnote{This set of $d$ vectors is ``overcomplete''; the vectors
$s_1^a, \dots, s_{d-1}^a$ are not linearly independent at
$\mathcal{I}^+$.} the subspace of the tangent space that is tangent
to $\mathcal{I}^+$.

Similarly, we obtain
\begin{equation}
W^a = u t^a + \Omega l^a  \, ,
\label{w2}
\end{equation}
where $u=t-r$ and $l^a$ is a smooth null vector field that is
nonvanishing at $\mathcal{I}^+$ but is not normal to $\mathcal{I}^+$.
Since $l^a$ is transverse to $\mathcal{I}^+$, it follows immediately
that the vector fields $t^a, l^a, s_1^a, \dots, s_{d-1}^a$ span the
full tangent space at each point of a neighborhood of $\mathcal{I}^+$.

Substituting eqs.~(\ref{K2}) and
(\ref{w2}) into eq.~(\ref{V}) to get the form of $V_i^a$ and
substituting this result into eq.~(\ref{chi0}), we see that $\chi$
takes the form
\begin{equation}
\chi = \Omega^{-1-d/2} [\Omega^2 F_1 + \Omega^3 F_2 + \Omega^4 F_3] \, .
\label{chi}
\end{equation}
Here $F_1$ denotes a linear combination of contractions of $\delta
\tilde{C}_{abcd}$ with the vectors $t^a, l^a, s_\mu^a$ that include
precisely two $t^a$'s, $F_2$ denotes a linear combination of these
contractions that include precisely one $t^a$, and $F_3$ denotes a
linear combination of these contractions that does not include any
$t^a$. The precise linear combinations occurring in $F_1$, $F_2$, and
$F_3$ depend, of course, upon the choice of $V_i^a$. By choosing
$V_1^a = V_3^a = t^a$ and taking each of $V_2^a$ and $V_4^a$ to be
either $W^a$ or a $K_\mu^a$, we obtain $F_2 = F_3 = 0$ and we can make
$F_1$ be any contraction of $\delta \tilde{C}_{abcd} t^a t^c$ with $l$'s
and $s_\mu$'s. This proves that the quantity
\begin{equation}
\Omega^{1-d/2} \delta \tilde{C}_{abcd} t^a t^c
\label{smooth1}
\end{equation}
must be smooth at $\mathcal{I}^+$. Given this result, we now may take
$V_1^a = t^a$ and choose $V_2^a, V_3^a, V_4^a$ to be any combinations of
$W^a$ and $K_\mu^a$ to conclude that
\begin{equation}
\Omega^{2-d/2} \delta \tilde{C}_{abcd} t^a
\label{smooth2}
\end{equation}
must be smooth at $\mathcal{I}^+$. Finally, given 
that expressions~\eqref{smooth1} and~\eqref{smooth2} are 
smooth at $\mathcal{I}^+$, if we then choose 
$V_1^a, V_2^a, V_3^a, V_4^a$ to be any combinations of
$W^a$ and $K_\mu^a$ we find that
\begin{equation}
\Omega^{3-d/2} \delta \tilde{C}_{abcd}
\label{smooth3}
\end{equation}
must be smooth at $\mathcal{I}^+$.

Since $t^a = \alpha_0 n^a - \Omega  \nablao  {}^a \alpha_0$, we
may restate the above result as follows: Let $\gamma_{ab}$ be an
arbitrary solution to the linearized Einstein equation (with smooth
initial data of compact support) off of $d$-dimensional Minkowski
spacetime for any $d>3$. Then the corresponding perturbed unphysical
Weyl tensor $\delta \tilde{C}_{abcd}$ in the Einstein static universe
is such that the three quantities
\begin{equation}
\Omega^{3-d/2} \delta \tilde{C}_{abcd}, \,\,  
\Omega^{2-d/2} \delta \tilde{C}_{abcd} n^a, \,\,
\Omega^{1-d/2} \delta \tilde{C}_{abcd} n^a n^c
\label{smooth}
\end{equation}
are all smooth at $\mathcal{I}^+$. Furthermore, since $\chi$ cannot
vanish identically for all $a_i^\mu$ and $b_i$ (unless
$\gamma_{ab}$ is pure gauge), it follows that at least one of the
above quantities must be nonvanishing somewhere on $\mathcal{I}^+$.

In $d=4$ dimensions, it turns out that $\delta \tilde{C}_{abcd}$
vanishes at $\mathcal{I}^+$, so, in particular, the first two terms in
eq.~(\ref{smooth}) vanish. The fact that, generically, in
$4$~dimensions all components of the perturbed unphysical Weyl tensor
are $O(\Omega)$ at $\mathcal{I}^+$ gives rise to the familiar
asymptotic ``peeling properties'' of the Weyl tensor in the physical
spacetime. However, the arguments leading to the vanishing of the first
two terms in eq.~(\ref{smooth}) (see \cite{g}) appear to be very special
to the $4$-dimensional case. We see no reason why these terms should vanish
when $d>4$. If so, the ``peeling behavior'' of the physical Weyl tensor for
$d>4$ would be qualitatively different from the $4$-dimensional case.

In any case, our above results establish that, 
$\Omega^{3-d/2} \delta \tilde{C}_{abcd}$ is smooth at $\mathcal{I}^+$,
and $\Omega^{1-d/2} \delta \tilde{C}_{abcd}$ cannot vanish at 
$\mathcal{I}^+$. This proves that in odd-dimensional spacetimes, the leading
order behavior of the perturbed unphysical Weyl tensor always begins with
a half-integral power of $\Omega$. Consequently, the unphysical metric fails
to be smooth at $\mathcal{I}^+$ (and $\mathcal{I}^-$) to the same degree
as it describes deviations of the physical metric from Minkowski spacetime.

It should be emphasized that our analysis does {\em not} show that it
is impossible to define a suitable notion of asymptotic flatness in
odd dimensions that admits a notion of Bondi mass and energy
flux. Indeed, it seems entirely possible that such a definition of
asymptotic flatness could be formulated in terms of an asymptotic expansion
of the physical metric in (integral and half-integral) powers of $1/r$,
analogous to the definition of asymptotic flatness in $4$-dimensions
originally given by Bondi et al \cite{b}. However, our analysis clearly
indicates that it will not be fruitful to seek a definition of asymptotic
flatness within the conformal null infinity framework for odd dimensional
spacetimes.

\medskip

We wish to thank Bob Geroch for suggesting that we directly
investigate the behavior of the perturbed Weyl tensor rather than the
perturbed metric for pertubations of Minkowski spacetime. This
research was supported in part by NSF grant PHY00-90138 to the
University of Chicago.


\begin{thebibliography}{99}

\bibitem{p} R. Penrose, Phys. Rev. Lett. {\bf 10}, 66 (1963).

\bibitem{f} H. Friedrich, ``Smoothness at null infinity and the
structure of initial data'', gr-qc/0304003.

\bibitem{cd} P.T. Chrusciel and E. Delay, Class. Quant. Grav. {\bf 19} 
L71 (2002); erratum Class. Quant. Grav. {\bf 19}, 3389 (2002).

\bibitem{hi} S. Hollands and A. Ishibashi, ``Asymptotic Flatness and
Bondi Energy in Higher Dimensional Gravity'', gr-qc/0304054;
``Asymptotic flatness at null infinity in higher dimensional
gravity'', hep-th/0311178.

\bibitem{gx} R. Geroch and B. Xanthopoulos, J. Math. Phys. {\bf 19},
714 (1978).

\bibitem{w} R.M. Wald, {\it General Relativity}, University of Chicago
Press (Chicago, 1984).

\bibitem{g} R. Geroch, in {\it Asymptotic Structure of Spacetime},
ed. by P. Esposito and L. Witten, Plenum Press (New York, 1977).

\bibitem{b} H. Bondi, M.G.J. van der Burg, and A.W.K. Metzner,
Proc. Roy. Soc. Lond. {\bf A269}, 21 (1962).

\end{thebibliography}
\end{document}